\documentclass[fleqn,10pt]{SelfArx}

\usepackage[english]{babel}
\usepackage{natbib}
\usepackage[table]{xcolor}
\usepackage{wrapfig}
\usepackage{tabularray}
\UseTblrLibrary{booktabs}
\usepackage{amsmath,amssymb}
\usepackage{graphicx}
\usepackage{hyperref}

\definecolor{color1}{RGB}{0,0,90}
\definecolor{color2}{RGB}{0,20,20}

\hypersetup{hidelinks,colorlinks,breaklinks=true,urlcolor=color2,citecolor=color1,linkcolor=color1,bookmarksopen=false,pdftitle={Hi-ReX SAG WG5 on Fundamental Physics and Cosmology},pdfauthor={Sara Turriziani et al.}}

\newcommand\araa{ARA\&A}%
\newcommand\apj{ApJ}%
\newcommand\apjl{ApJL}%
\newcommand\apss{Ap\&SS}%
\newcommand\aap{A\&A}%
\newcommand\mnras{MNRAS}%
\newcommand\nat{Nature}%

\JournalInfo{White paper on High-Resolution X-ray Imaging}
\PaperTitle{Advancing Fundamental Physics and Cosmology with high-resolution X-ray imaging}
\Archive{}

\Authors{Sara Turriziani \textsuperscript{1}*, Herman L. Marshall\textsuperscript{2}, Daryl Haggard\textsuperscript{3}, Scott Randall\textsuperscript{4}, Mayura Balakrishnan\textsuperscript{3}, Tom Maccarone\textsuperscript{5}, Nicole Ford\textsuperscript{3}}
\affiliation{\textsuperscript{1}\textit{Centro de Astronomía UA}} 
\affiliation{\textsuperscript{2}\textit{Massachusetts Institute of Technology}} 
\affiliation{\textsuperscript{3}\textit{McGill University and the Trottier Space Institute}} 
\affiliation{\textsuperscript{4}\textit{Center for Astrophysics $\vert$ Harvard \& Smithsonian}} 
\affiliation{\textsuperscript{5}\textit{Texas Tech University}} 

\affiliation{*\textbf{Corresponding author}: Sara.Turriziani@uantof.cl} 
\Keywords{X-ray astronomy --- General Relativity --- Cosmology --- Intracluster medium --- Dark Matter --- Gravitational waves}
\newcommand{\keywordname}{Keywords}

\Abstract{Black Holes are the key to solving many unanswered questions in fundamental physics: in particular, the very extreme properties shown by supermassive black holes at the centers of galaxies make them obvious candidates for testing gravity theories in the strong-field regime. Since X-rays are generated by matter under extreme physical conditions, ultra-high resolution X-ray imaging (uXRI) will directly image the region near the event horizon of black holes in X-rays, similar to the Event Horizon Telescope in the radio band, enabling unprecedented tests of General Relativity and alternative theories of gravity near supermassive black holes. 
On the other hand, clusters of galaxies hold the potential of unveiling many unknowns in cosmology. uXRI will unlock this potential by probing small-scale plasma properties in the intracluster medium, providing the missing link required to establish galaxy clusters as reliable tools for high-precision cosmology. Moreover, uXRI will enable mapping of Dark Matter from galaxy cluster dynamics via proper motion measurements.  
Finally, uXRI will open a new field of precision X-ray astrometry, allowing for measuring pulsar parallaxes to support nanoHertz gravitational wave searches.}

\begin{document}

\flushbottom
\maketitle
\thispagestyle{empty}

\section{Enabling New Discoveries with High Angular Resolution}

\begin{figure*}[h]
    \centering
    \includegraphics[width=0.8\textwidth]{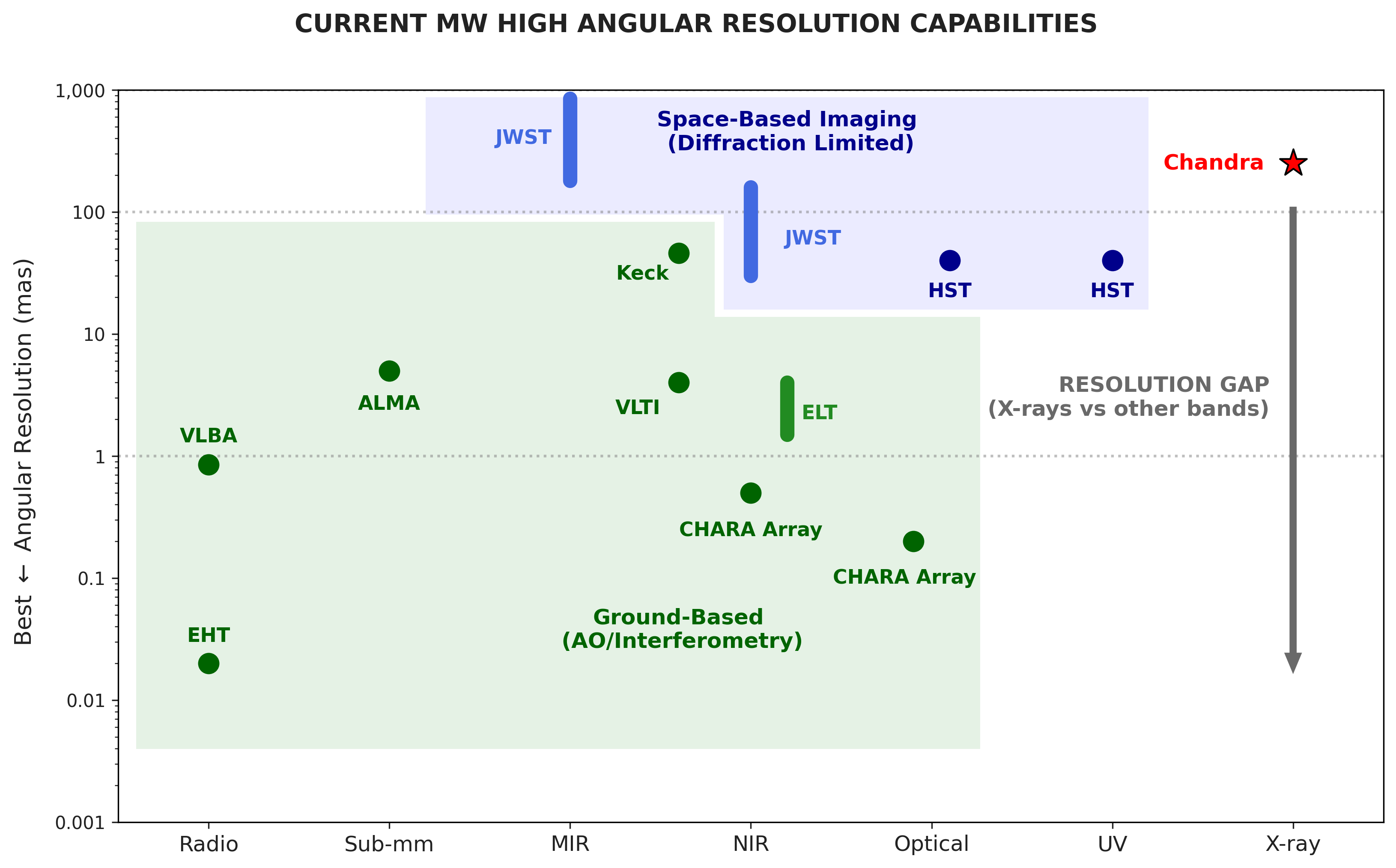}
    \caption{\textit{The advance in technologies such as interferometry and adaptive optics enabled mas angular resolution observations from ground in the radio (VLBA), sub-mm (ALMA), and NIR/MIR bands, with the EHT pushing this limit down to $20\ \mu\text{as}$ in radio. Current X-ray angular resolution capabilities instead approach those of HST and JWST. However, they are not sufficient to match the angular resolution delivered by interferometry and adaptive optics based ground facilities.}}
    \label{fig:MW_cap}
\end{figure*}

As shown in Fig. \ref{fig:MW_cap}, current X-ray angular resolution limits approach those of HST and JWST. \textit{Chandra}'s nominal angular resolution ($0.5-1''$) can be pushed down to $0.2-0.3''$, using advanced techniques such as Energy-Dependent Subpixel Event Repositioning \citep[EDSER,][]{Li2004}. This capability allows, for example, to measure binary AGN at $\sim 100-150$ pc scales \citep[e.g.][]{Fabbiano2011,TriF2024}, with the possibility to test further the presence of multiple sources using Bayesian estimators like BAYMAX \citep{Foord2019H}.

However, even these enhanced X-ray capabilities are completely insufficient to match the high angular resolution delivered by present ground-based interferometry and adaptive optics facilities. In fact, X-ray astronomy faces a huge resolution gap with respect to ground observatories, which are capable of achieving mas angular resolution in the radio, sub-mm, and NIR/MIR bands, with the Event Horizon Telescope (EHT) pushing this limit further down to $20\ \mu\text{as}$ in radio/sub-mm. 

Next-generation, ultra-high angular resolution X-ray imaging (uXRI, also referred to in this context as HiReX) will finally close this gap and transform our understanding of cosmology and fundamental physics \citep{Gen-astro2010,2021ExA....51.1081U}. It will enable more rigorous testing of General Relativity (GR) near event horizons through direct imaging of the immediate surroundings of supermassive black holes, along with high-resolution reverberation mapping \citep{2021SPIE11444E..1EU}. Such studies will complement the recent advancements by the EHT. 

In addition to fundamental physics, ultra-high angular resolution will revolutionize our understanding of cosmology by probing the microphysics of the intracluster medium (ICM). By resolving the fine structures of cluster shocks and cold fronts, these observations will transform ICM studies from one-dimensional edge profiles into spatially resolved, multi-dimensional plasma diagnostics. This will allow us to directly constrain, for the first time, the transport processes that influence turbulent dissipation and clusters' evolution over cosmic time. Moreover, uXRI studies of galaxy cluster dynamics will open a new window on the studies of Dark Matter (DM): in fact, measuring proper motions for galaxies in nearby clusters to high accuracy will allow for mapping DM, and also to detect the predicted dynamical friction from DM in local groups.

Finally, we highlight uXRI's potential for X-ray pulsar astrometry: by measuring more accurate parallax distances to nearby pulsars, uXRI will revolutionize pulsar timing searches for gravitational waves.

\section{Key Science Cases}

 In the following, we will detail each science goal, addressing two key questions: (1) why high angular resolution is needed, and (2) what specific resolution is required. Additionally, we define science requirements for energy bands, flux levels, and effective area ($A_{\text{eff}}$), and discuss considerations for added capabilities, such as polarimetry. Table \ref{tab:science_summary} provides a high-level summary of these science cases and observables, which have been integrated into the Hi-ReX SAG document \citep{HiReX-SAG_report}.

\begin{table*}[htbp]
\centering
\caption{Science Case Summary}
\label{tab:science_summary}

\begin{tblr}{
  width = \textwidth,
  colspec = {
    X[2.2,l] 
    X[2.2,l] 
    X[1.5,l] 
    X[3.5,l] 
    X[0.8,c] 
  },  
  row{2-4} = {gray!20},
  row{5}   = {white},
  row{6}   = {gray!20},
  row{1}   = {font=\bfseries, bg=white},
  cells    = {valign=m},
}
\toprule
Topic & Science Case & Resolution & Other Requirements & Section \\
\midrule
 & Sgr A* flares & $10\,\mu\text{as}$ & 2--10 keV flux, Chandra-like $A_{\text{eff}}$, deep obs & 2.1.1 \\
Testing General Relativity & Fe K$\alpha$ (RQ AGN) & $< 1\,\mu\text{as}$ & 6--7 keV flux, Chandra-like $A_{\text{eff}}$, ~~~~~deep obs & 2.1.2 \\
 & BH shadow (RQ AGN) & $0.1\text{--}10\,\mu\text{as}$ & 2--10 keV flux, Chandra-like $A_{\text{eff}}$, deep obs & 2.1.3 \\
\midrule
Plasma Physics & ICM microphysics & $50\text{--}200\,\text{mas}$ & 0.5--7 keV, $10\text{--}100\times\text{Chandra } A_{\text{eff}}$ & 2.2 \\
\midrule
Galaxy cluster dynamics & Dark matter distribution, cluster assembly & $\mu\text{as/yr}$ & flux limit $\sim 10^{-15}\,\text{erg s}^{-1}\text{cm}^{-2}$ & 2.3 \\
\midrule
Gravitational Waves & Identify sources of nHz waves & $\mu\text{as}$ & flux limit $\sim 10^{-15}\,\text{erg s}^{-1}\text{cm}^{-2}$ & 2.4 \\
\bottomrule
\end{tblr}
\end{table*}

\subsection{Testing alternative theories of GR in X-ray}

Since X-rays are generated by matter under extreme physical conditions, such as the vicinity of compact objects or in the centers of AGN, they allow us to study general relativity (GR) in the strong-field regime. An uXRI mission would open a new window in the testing of GR and alternative gravity theories in the extreme gravitational field near black holes, offering direct evidence of spacetime warping and accretion inflows and outflows.

Electromagnetic tests of GR are complementary to studies performed with gravitational waves (GW). In fact, GW probe the ``dynamic'' spacetime in binary systems during the coalescence phase, as they are sensitive to the field equations of the theory of gravity. On the other hand, electromagnetic tests are sensitive to the motion of particles and can be used to get information on the interaction between matter and gravity in ``static'' spacetime (e.g. Sgr A*), studying phenomena such as the bending of light and gravitational redshift \citep{Bambi2017, 2022hxga.book...81A}.

\subsubsection{Sgr A* flares}\label{subsubsec:flares}

Sagittarius A* (Sgr A*; $M = 4 \times 10^6  M_{\odot}$, $d = \sim 8$ kpc) is our neighbour supermassive black hole (SMBH). It is considered to be a dormant black hole, accreting at a very low rate through a radiatively inefficient accretion flow.

Sgr A* has been monitored since the 1990s and it shows flaring activity in the radio, sub-millimiter, infrared, and X-ray bands \citep[e.g.][]{Dodds-Eden2011,Witzel2021,2025Fellenberg}. Simultaneous multi-wavelength campaigns have shed light on the underlying physical mechanism originating these flares, with current modeling indicating that they can be explained with synchrotron or synchrotron-self Compton emission from non-thermal electrons \citep{Dodds-Eden2009,Ponti2017,GRAVITY2021}. 

These flares are stochastic and follow a power-law distribution, which implies that the brightest flares are extremely rare. Data from 25 years of \textit{Chandra} monitoring ($\sim 7$ Ms) permit a complete characterization of flare demographics in the X-rays: \cite{Sumners2026} found 100 flares in total, among which only 5 were very bright. Moreover, their analysis shows that many, but not all, of the moderate flares have at least two peaks in their X-ray light curve, and all strong flares have a substructure. 

GRMHD-modeling predicts these flares to arise from magnetic reconnection events in the accretion disc, and emission is expected to be well-localized with high surface brightness over a small area in the case of bright flares \citep{Ripperda2020,Chatterjee2021}. Current observations also favor the idea of a few monolithic structures rather than hundreds of tiny emission sites in the case of bright flares: in fact, GRAVITY have seen centroid shifts \citep{GRAVITY2018, GRAVITY2020B}, which suggest orbiting hotspots at 6-10 gravitational radii ($R_g$). Sgr A* is therefore the best laboratory in the local Universe to study magnetic reconnection, and general relativity at work. Unfortunately, X-ray observations at present cannot perform a similar tracking, given the gap in resolution with respect to ground facilities at other wavelengths (see Fig. \ref{fig:MW_cap}).

A $10\mu\text{as}$ uXRI can detect hotspots orbiting Sgr A* at $\ge 6 R_g$. Since 6$R_g$ ($\sim 30 \mu\text{as}$) is the ISCO location in the Schwarzschild metric, this will allow us to measure the movement of material in the inner part of the disc, and compare with the predictions of GR and alternative theories of gravity \citep{Shahz2022}. While the GRAVITY interferometer is seeing hints of this in the infrared, in X-rays we will measure the hottest, most energetic electrons, closer to the SMBH. It is important to track hotspots along complete orbits to verify whether the hotspot centroid returns to its starting point, proving if the orbit is closed. From 25 years of Chandra observations, it is possible to infer the rate for flares lasting long enough to complete at least one full ISCO orbit (30 minutes for Sgr A*), and use it to estimate the monitoring time required to catch a few of these flares. We estimate that a monitoring time in the range 250-500 ks will be necessary for an uXRI with Chandra-like effective area. The prolonged observations of Sgr A* needed to catch bright X-ray flares will also provide data during the SMBH's quiescent periods. This will enable deep statistical studies of the persistent X-ray emission generated by the surrounding Bondi accretion region.

\begin{figure*}[h]
    \centering
    \includegraphics[width=0.8\textwidth]{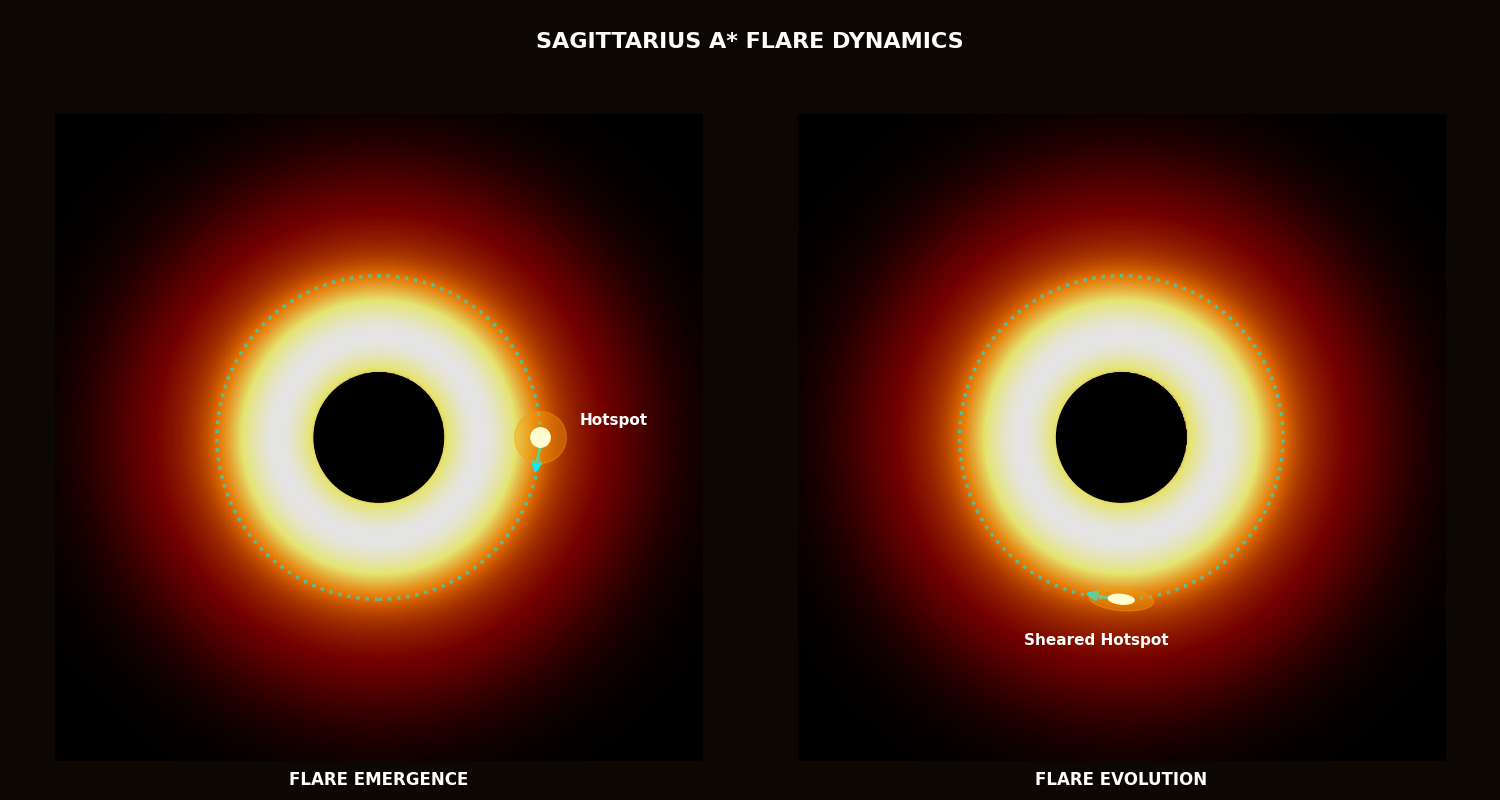}
    \caption{\textit{Schematic representation of an orbiting hotspot near Sgr A*. \textbf{Left panel ($T=0$)}: An idealized, spherical hotspot emerges in the inner accretion disk in connection of an X-ray flare. \textbf{Right panel}: The hotspot advances along its orbit (by a $90^\circ$ clockwise rotation), with elongation and morphological distortion induced by intense gravitational shear near the event horizon.}}
    \label{fig:hotspot}
\end{figure*}

An uXRI with resolution of $10 \mu\text{as}$ can therefore record a literal movie of a magnetic reconnection event accelerating plasma while simultaneously plunging into the event horizon (Figure \ref{fig:hotspot}). In fact, it can detect a hotspot orbiting at the ISCO and track its evolution, measuring the hotspot centroid with a few $\mu\text{as}$ precision along its orbit. An uXRI with resolution of $1 \mu\text{as}$ will push the centroid accuracy down to $< 1 \mu\text{as}$.  
These estimations assume a Chandra-like effective area. It would be useful to also add polarimetric capabilities, to compare with GRAVITY in the NIR \citep{GRAVITY2023, Levis2024}. We foresee promising synergies with ground based optical and radio facilities, such as ALMA/ALMA2040, EHT/ngEHT, ELTs, VLTI (GRAVITY/GRAVITY+).

\subsubsection{X-ray Reverberation Mapping of Radio Quiet AGN}\label{subsubsec:reverberation}

The Fe K$\alpha$ line is by far the strongest X-ray line of any element in the reflection spectrum of AGN, and it provides information on the dynamics and thermodynamics of the reprocessing material. The width of the Fe K$\alpha$ is used to pinpoint the emission region and to discriminate the origin of the line, with the narrow Fe K$\alpha$ lines interpreted as fluorescence lines from cold material away from the inner accretion disk.
This line is intrinsically narrow ($\sim 3.5$~eV); however, the presence of the SMBH introduces relativistic effects that are stronger the closer to the black hole: the combination of relativistic Doppler effect and gravitational redshift results in a highly broadened and skewed line profile. Such broad lines have been detected in many radio-quiet  (RQ) AGN, especially in bright nearby AGN. The emission region of this broad Fe K$\alpha$ line is expected to arise in the accretion disc at $\sim 10-30 R_g$.
The strong dependency of the line profile with spin and inclination angle of the disc makes it the perfect tool to probe the relativistic effects due to the proximity of the SMBH, the mode of BH accretion, and the unification theory of AGN.

To map fundamental spacetime geometry and detect e.g. the wobble predicted by the Lense-Thirring effect, we need to track the shifting centroids of line emission regions in the inner accretion disk. This requires a high-framerate, spatially resolved sub-microarcsecond astrometry. 
By dividing deep observations into shorter temporal frames, it is possible to measure the phase-dependent shift of the line's centroid, directly tracking the kinematics of the accretion disk at distances of approximately 10 $R_G$ from the central engine. Although an imaging narrow band centered on the broad fluorescent Fe K$\alpha$ around 6 keV technically should permit separation of the accretion disk and relativistic jet emission components, disentangling the two may prove difficult in radio-loud systems.
For this reason, we prioritize radio-quiet AGN as good targets for line centroid measurements: these probe the ``pure'' strong gravitational field and test the Kerr metric without strong jet contamination. To this end, we consider the flux-limited sample from the FERO survey \citep{FERO2010}, complemented with the black hole masses measured from reverberation mapping campaigns, finding 14-17 promising bright targets.  
Sub-$\mu\text{as}$ ($\sim 0.1\mu\text{as}$) resolution is needed to measure the line centroid. This can be achieved for our target sources with a $\sim 1 \mu\text{as}$ uXRI with a tight PSF. We assume an uXRI instrument with Chandra-like effective area at 6-7 keV, and the capability for monitoring the targets with deep exposures.

\subsubsection{Direct imaging BH shadow}

An SMBH with an optically thin accretion inflow should display a shadow surrounded by a ring of emission \citep[e.g.,][]{Falcke+2000a}. The angular size of this shadow is comparable to the ISCO, scaling linearly with BH mass and inversely with distance. In the sub-mm regime, the EHT has directly imaged two of the largest known BH shadows, corresponding to Sgr A* and M87* \citep{eht2019:m87i,eht:2022sgrai}. 
M87* is nearly $10^{10} M_{\odot}$ at a distance of 15.8 Mpc, and its angular shadow size is 42$\mu\text{as}$. Beyond M87* and Sgr A*, the next largest BH shadows have predicted angular sizes ranging from sub-$\mu\text{as}$ to, at most, tens of $\mu\text{as}$ \citep{ETHER2023}. 
 
The next generation of resolvable BH shadows consists of a mixture of radio-loud and radio-quiet AGN. The  radio-loud, low luminosity AGN  ---for example, NGC 4594 and M84--- at $z\sim0$ tend to have larger predicted SMBH shadows \citep{2025Zhang}. These targets, however, typically require longer exposures and/or stacked monitoring observations due to their relatively faint X-ray flux ($\lesssim1\times10^{-12}$ erg~cm$^{-2}$~s$^{-1}$; see \citealt{2016Connolly,2023Bambic,2025Ford}). Meanwhile, based on reverberation mapping and X-ray luminosity, we estimate that the angular diameter of the ISCO region for almost all standard radio-quiet AGN is sub-microarcsecond: e.g., NGC 4151 has an estimated photon ring with a diameter of just 0.2 $\mu\text{as}$. 

The required instrument resolution depends on which BH shadow targets are prioritized. To image the radio-quiet, X-ray bright nearby AGN, we recommend a nominal instrument capability of at least a tenth of a microarcsecond ($0.1 \mu\text{as}$). For low luminosity AGN, the resolution requirement is slightly less strict; $1-10 \mu\text{as}$ may be sufficient to resolve some of the largest SMBH shadows. However, the finer the resolution, the larger the sample of SMBHs available for direct imaging. We assume a Chandra-like effective area at 6-7 keV. For low luminosity AGN, moderate to deep exposures ($\gtrsim50-100$ ks) or monitoring may be needed to detect the photon emission ring. Such an X-ray ``Event Horizon Telescope'' will complement EHT/ngEHT observations, allowing for multi-frequency modeling of black hole shadows. 

\subsection{ICM Plasma Microphysics}

\begin{figure*}[h]
    \centering
    \includegraphics[width=0.8\textwidth]{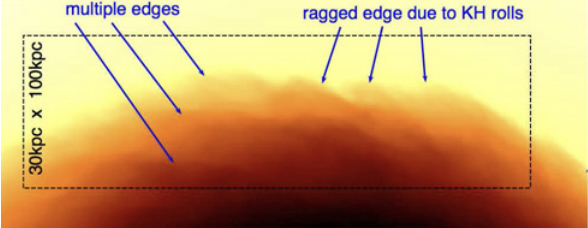}
    \caption{\textit{Synthetic X-ray image based on simulations of sloshing cold fronts in galaxy clusters. For this case of a relatively low-viscosity ICM Kelvin-Helmholtz instabilities generate sub-kpc features along the cold front edge. A higher ICM viscosity gives smoother edges. Therefore, resolving the structure of these edges places limits on the microphysics of the ICM plasma. Figure adapted from \cite{2013ApJ...764...60R}.}}
    \label{fig:hki}
\end{figure*}

The intracluster medium (ICM) is a weakly collisional, high-$\beta$ plasma whose properties rule the thermodynamic evolution of galaxy clusters. Highly localized transport processes determine how the energy from AGN-driven feedback and hierarchical mergers is dissipated and thermalized across cosmic time \citep{Fabian2012, McNamara2012}. At present, we are lacking strong observational constraints on these crucial processes, and this introduces degeneracies and biases in cosmological hydrodynamical simulations, see e.g., BAHAMAS \citep{McCarthy2017}, Rhapsody-C \citep{Pellissier2023}, and FLAMINGO \citep{Schaye2023}. 
Ultra-high angular resolution X-ray imaging is the key to finally resolve these bottlenecks, as it will permit to measure spatially resolved diagnostics of the ICM. In the following subsections, we focus on the most promising observational sites to directly measure these quantities: contact discontinuities (Section \ref{contactdis}) and shock fronts (Section \ref{shocks}). Ultimately, resolving these small-scale plasma interfaces is the missing link required to properly use galaxy clusters for high-precision cosmology.

\subsubsection{Contact discontinuities}
\label{contactdis}
Despite decades of study with Chandra and other major X-ray missions, the microphysics of the ICM remains, in many respects, poorly understood \citep{ZuHone2016}. Ultra-high angular resolution X-ray imaging would open a new observational regime for intracluster plasma physics by resolving the narrow interfaces where the weakly collisional ICM reveals its effective transport properties, e.g. at ``contact discontinuities''. Cold fronts, AGN-inflated bubble rims, multiphase filaments, and ram-pressure-stripped galaxy tails place the hot ICM in direct contact with cooler gas. The intrinsic widths, sharpness, and smoothness of these boundaries encode the suppression of thermal conduction and particle diffusion, the effective viscosity (Figure~\ref{fig:hki}), and the role of magnetic draping \citep{Vikhlinin2001,MarkevitchVikhlinin2007,2013ApJ...764...60R,WangMarkevitch2018}. 
These transport processes determine how long entropy, temperature, metallicity, and density substructure survive in the ICM, affecting the thermodynamic profiles and intrinsic scatter of cluster observables used in cosmological analyses.
With Chandra, such fronts are already known to be sharper and smoother than simple hydrodynamic expectations, but current observations generally provide only a small number of resolution elements across the relevant structures \citep{Vikhlinin2001,Werner2016,ZuHone2016}. An uXRI mission would allow front widths and Kelvin–Helmholtz cutoff scales to be measured locally along many independent interface segments, converting qualitative evidence for suppressed transport into spatially resolved constraints on ICM microphysics.

Existing Chandra observations show that cold fronts are often unresolved at kpc scales and narrower than Coulomb mean-free-path expectations, demonstrating suppressed transport but not measuring the true transport layer. A next-generation mission should therefore resolve $\sim$30–300 pc scales in the nearest bright systems, corresponding to $\sim$0.05''–0.2'' imaging for low-redshift clusters like Virgo and Perseus. This would allow direct measurements of interface widths, small-scale Kelvin-Helmholtz instability (KHI) suppression (Figure~\ref{fig:hki}), magnetic draping, and mixing layers, rather than simply placing kpc-scale upper limits on conduction and diffusion. Mapping detailed structure at the resolution limit requires $\gtrsim 100$ counts, while detailed thermodynamic structure requires $\gtrsim 1000$ counts. Therefore, fully characterizing an interface layer both along and across the layer requires $\gtrsim 10^5$ counts.

\subsubsection{Shock fronts}
\label{shocks}

Merger shocks and AGN-driven weak shocks provide direct laboratories for understanding how energy is thermalized in the ICM. In mergers, shocks convert gravitationally driven bulk motions into heat, turbulence, and nonthermal particle populations \citep{MarkevitchVikhlinin2007,Russell2022,vanWeeren2019}; in cool cores, repeated weak shocks and sound waves generated by AGN outbursts may distribute jet power through the cluster atmosphere \citep{Fabian2003,Fabian2006,Forman2007,Randall2011,Randall2015}. The structure of these fronts therefore encodes fundamental plasma physics: how rapidly electrons are heated relative to ions, how efficiently conduction smooths temperature gradients, whether viscosity damps corrugations and turbulence, and whether cosmic-ray pressure modifies the shock jump conditions. These questions also feed directly into cosmology, because merger-driven turbulence, bulk motions, and non-thermal pressure support are major sources of scatter and bias in cluster mass estimates and in the mapping between dark-matter halo growth and observable ICM properties.

Ultra-high-resolution X-ray imaging would turn shock studies from one-dimensional edge fitting into spatially resolved plasma diagnostics. Rather than treating shocks as idealized spherical or planar discontinuities, such observations would map their local curvature, intrinsic or projected width, ripple/corrugation spectrum, and sector-to-sector Mach-number variations. This would distinguish true physical broadening from projection and geometric effects, use shock-front corrugation as a tracer of upstream turbulence, and test whether weak shocks, sound waves, turbulent mixing, or conduction dominate the dissipation of AGN feedback energy. In nearby cool-core clusters, resolving shocks together with cavities, rims, filaments, and contact discontinuities would directly measure how jet energy couples to the surrounding ICM \citep{Fabian2003,Birzan2004,McNamara2007,Randall2015}.
Such measurements would provide the microphysical closure needed to model how feedback and mergers set cluster pressure profiles, thermal histories, and the level of non-thermal support in cosmological simulations.

Chandra observations of nearby cool cores such as Perseus and M87 reveal weak shocks, ripples, cavities, rims, and filaments on $\sim$kpc and sub-kpc scales \citep{Fabian2003,Fabian2006,Forman2007}, while deep observations of merger shocks such as Abell 2146 measure projected shock widths of order $\sim$10–20 kpc across several-hundred-kpc fronts \citep{Russell2022}. Resolving these structures requires imaging substantially finer than the front width: $\sim$0.1'' resolution to study sub-kpc AGN-shock structure in the nearest systems, and $\lesssim$0.5''–1'' resolution to map kpc-scale merger-shock widths, curvature, and corrugation in low-redshift clusters. A goal of 0.1'' imaging would enable local Mach-number maps, shock-width measurements, and ripple/corrugation spectra rather than azimuthally averaged edge profiles. Mapping detailed structure at the resolution limit requires $\gtrsim 100$ counts, while detailed thermodynamic structure requires $\gtrsim 1000$ counts. Therefore, fully characterizing a shock front both along and across the front requires $\gtrsim 10^5$ counts.

\subsection{Galaxy cluster dynamics}

Understanding the dark matter properties of nearby galaxies and clusters would be vastly improved with full phase space information. Obtaining three of the six dimensions of position-velocity phase space -- radial velocity and the two components of position on the plane of the sky -- is straightforward now. Distance determinations to high enough precision to provide useful localizations of galaxies within clusters may be obtained by LISA through standard siren distances in the best case, or through surface brightness fluctuations more routinely. The other two dimensions, the transverse velocities, require astrometric capabilities that exceed current instrumentation, but that might be achievable with the Next Generation Very Large Array \citep{2018ASPC..517..663M}.

Proper motions of galaxies offer a few other key capacities for understanding cosmology. For example, proper motions of AGN in the Bullet Cluster can reveal the relative masses of the two colliding clusters \citep{2018ASPC..517..663M}, and relative proper motions of galaxies in groups can reveal whether the dynamical friction expected from dark matter models, but not expected for Modified Newtonian Dynamics models, is taking place \citep{2017MNRAS.467..273O}. Simply getting the directions of proper motions -- toward or away from the center of the clusters -- is sufficient for determining if spiral galaxies in clusters are disproportionately on their first approaches through the galaxy clusters \citep{2018ASPC..517..663M}.

Proper motions with accuracies of about $1~\mu\text{as}$/year would be enough for galaxies in nearby clusters (e.g. Virgo and Fornax) from either AGN (ideally) or ensembles of bright X-ray binaries, as $1~\mu\text{as}$/year at 16~Mpc corresponds to a velocity of 75 km~s$^{-1}$. A flux sensitivity of about $10^{-15}$ erg~cm$^{-2}$~s$^{-1}$ (corresponding to a luminosity of about $3\times10^{38}$ erg~cm$^{-2}$~s$^{-1}$) would be sufficient to ensure that nearly all large Virgo galaxies can have their proper motions measured. Many have AGN at this level, so that a single object can be used to estimate the galaxy's center of mass motion. Most of the rest will have at least a dozen X-ray binaries at this luminosity; for these each individual object will have a motion relative to the center of mass of the galaxy of order the galaxy's velocity dispersion. For a velocity dispersion of 400 km~s$^{-1}$, 12 X-ray binaries will then yield a mean velocity with accuracy of about 120 km~s$^{-1}$; many large Virgo galaxies have more than this number of X-ray binaries.

To test for dynamical friction from dark matter in local groups, the problem is somewhat easier. \cite{2017MNRAS.467..273O} find that the expected signal is about $2~\mu\text{as}$/year for the M81 group, at about 4~Mpc distance. This corresponds to about 35 km~s$^{-1}$. Because the galaxies are smaller than in the Virgo Cluster, a similar number of X-ray binaries can be used for the galaxies that do not have AGN.

\subsection{Pulsar parallaxes and searches for nanoHertz gravitational waves}

Many pulsars used in pulsar timing arrays \citep{2025Ap&SS.370..124T} are also X-ray sources within 1 kpc of Earth. An uXRI with $\mu\text{as}$ spatial resolution can measure accurate parallaxes that constrain distances to a fraction of a gravitational wavelength. Obtaining even a handful of pulsars with sufficiently accurate distances would revolutionize pulsar timing searches for gravitational waves \citep{2026arXiv260310120T,2026ChPhL..43f1102W}.

At the present time, pulsar timing arrays are sensitive only to the ``Earth term'' of gravitational radiation -- the effects of the gravitational waves passing near the Earth that leave residuals with a characteristic correlation pattern on the sky (strong positive correlations for pulsars with angular separations near 0$^\circ$ or 180$^\circ$, and anticorrelations for separations near 90$^\circ$ -- \citealt{1983ApJ...265L..39H}). If the pulsar distances are known precisely, then their source terms -- the timing residuals due to gravitational waves passing near the pulsars -- can be used as well. This modestly increases the signal-to-noise for stochastic backgrounds, but substantially increases the signal-to-noise and localization capabilities for individual sources of gravitational waves.

With distance estimates to 0.1 pc, the correlations between the pulsar timing residuals can then be used to localize individual sources of nHz gravitational waves, rather than just the stochastic background \citep{2011MNRAS.414.3251L,2026arXiv260310120T, 2026ChPhL..43f1102W}. With precise pulsar distance, localizations of individual sources detected with pulsar timing arrays will go from hundreds of square degrees to sub-degree, opening up the possibility of secure localizations, and hence the ability to do standard siren cosmology with nanohertz gravitational waves. Furthermore, the waves from a single source can then be used to solve for the distances to other pulsars in the timing array, in a manner similar to how fringe-finding measurements with the Very Long Baseline Array are used to measure plate tectonic drift. With distance solutions for those pulsars, their use as probes of the interstellar medium, and of substructure in the Milky Way's dark matter potential can be refined.

As the parallax distances need to be measured to about 0.1 parsec, the parallax shifts need to be accurate to 1 part in 1000 for sources at 100 pc, and the error budget will scale inversely with the square of the  distance -- i.e. the precision must be $10~\mu\text{as}$ $\left(\frac{d}{100 {\rm pc}}\right)^{-2}$. As this is a centroiding measurement, the precision can be obtained through a combination of sensitivity and resolution. Some of the best cases, like PSR~J0030+0451 are well timed in NanoGRAV, and also relatively bright in X-rays, but it may require going out to $\sim$kpc distances to obtain a good sample of well-timed, X-ray bright ($F_X>10^{-13}$~erg~cm$^{-2}$~s$^{-1}$) pulsars \citep{2021cosp...43E1186R}. Tolerating pulsars a factor of 10 or so fainter in X-rays opens up a much larger sample of objects. Assuming a reasonable sensitivity level of $10^{-15}$~erg~cm$^{-2}$~s$^{-1}$, so that objects at $10^{-14}$~erg~cm$^{-2}$~s$^{-1}$ will be at $\sim7$ sigma, angular resolution of about $1~\mu\text{as}$ would be needed for this work.

Importantly, it is likely that the level of precision needed for this work cannot be achieved in the radio band from the ground, because of systematic limits in how well the effects of the ionosphere and troposphere can be corrected. Present telescopes are limited to precision of about $10~\mu\text{as}$ \citep{2014ARA&A..52..339R}, and while some substantial improvement will likely result from higher sensitivities with the ngVLA, as fainter calibrators, nearer to the sources, can be used, it is unlikely that sub-$\mu\text{as}$ precision will be possible in the foreseeable future.


\end{document}